\documentclass[%
 reprint,
 amsmath,amssymb,
 aps,
prb,
floatfix,
]{revtex4-2}

\usepackage{graphicx}
\usepackage{xcolor}

\usepackage{amsmath}
\usepackage{mathtools}
\usepackage{natbib}
\usepackage{dcolumn}
\usepackage{bm}
\usepackage{float}
\usepackage{physics}

\begin{document}

\preprint{APS/123-QED}

\title{Dephasing-induced jumps in non-Hermitian disordered lattices}

\author{Emmanouil T. Kokkinakis$^{1,2}$}

\author{Konstantinos G. Makris$^{1,2}$}%

\author{Eleftherios N. Economou$^{1,2}$}
\affiliation{$^1$
  Department of Physics, University of Crete, 70013 Heraklion, Greece
}%
\affiliation{$^2$
 Institute of Electronic Structure and Laser (IESL), FORTH, 71110 Heraklion, Greece
}%
\date{\today}

\begin{abstract} Changes in the wave function's phase during propagation in a random Hermitian lattice, a process known as dephasing, results in diffusion rather than Anderson localization. However, when non-Hermiticity is introduced, the wave behavior changes drastically. In particular, we demonstrate that in weakly disordered non-Hermitian lattices, dephasing enhances eigenmode localization, which results in abrupt jumps between spatially distant regions. These jumps, which are absent under purely coherent conditions, emerge from the interplay between complex disorder and dephasing.
\end{abstract}

\maketitle


\section{INTRODUCTION}

In recent years, photonics has been established as the most successful platform for exploring non-Hermitian systems \cite{christodoulides_2003}, due to advanced experimental techniques that enable precise manipulation of non-Hermiticity \cite{wiersma_1997, lagendijk,schwartz_2007,lahini_2008}. This is in contrast to most other areas of physics, where controlling openness poses substantial practical challenges. The introduction of the concepts of parity-time ($\mathcal{P}\mathcal{T}$) symmetry \cite{bender_1998, bender_1999, bender_2002, makris_2008, el_ganainy_2007, Musslimani_2008, guo_2009, ruter_2010, makris_2010, bender_2007} and exceptional-points \cite{moiseyev_2011, berry_2004, heiss_2004, wiersig_2008, lee_2009, miri_2019} in optics has led to intense research activity in the emerging field of non-Hermitian photonics over the past decade \cite{el_ganainy_2018, regensburger_2012, feng_2013, zhang_2018, hodaei_2014, konotop_2016, feng_2017, pile_2017, gbur_2018, ozdemir_2019, longhi_2018, chong_2011, ge_2012, wimmer_2015, szameit_2011, zuener_2015, peng_2014, feng_2014, peng_2014_2, assawaworrarit_2017, zhang_2018, makris_2015, makris_2016, brandstotter_2019, longhi_2019, komis_2024}. The main approach for realizing non-Hermitian Hamiltonians involves designing structures with spatially distributed loss and/or gain materials. In semiclassical optics, such dissipative and active materials, are easily accessible, unlike in condensed matter or quantum systems. In addition to systems with on-site gain/loss potentials, extensive theoretical and experimental research has focused on non-Hermitian lattices with asymmetric couplings \cite{hatano_1996, hatano_1997, jiang_2019, liu_2022, liu_2021, gao_2023, yao_2018, gong_2018, lee_2016, harari_2018, bandres_2018, okuma_2020, li_2020, longhi_2021_1, ezawa_2022, faugno_2022, zhu_2022, weidemann_2020, komis_2023, longhi_2021, chen_2023, gandhi_2023, xiao_2024, kokkinakis_2024, borgnia_2020}.

In contrast to these relatively recent advances, Anderson localization is a phenomenon that has been extensively studied for over six decades \cite{Akkermans, anderson_2}. As first introduced by Anderson in his seminal work \cite{anderson}, it predicts the absence of diffusion of waves propagating through lattices with random on-site disorder. Although initially rooted in the context of solid-state physics, Anderson localization has since had significant impact over a wide range of fields, including disordered photonics, Bose-Einstein condensates, imaging, and acoustics \cite{segev_2013, wiersma_2013, damski_2003, condat_1987}. In these areas, direct experimental investigations are not hindered by many-body interactions or temperature-dependent effects, allowing for direct observations of localization phenomena. With notable exceptions such as random lasers \cite{wiersma_2008,cao,vanneste}, most studies have focused on conservative systems governed by Hermitian Hamiltonians. 

Recently, attention has been drawn to non-Hermitian disordered problems \cite{makris_2017, makris_2020,tzortzakakis_2020,huang_2020,tzortzakakis_2020_2, Sukhachov_2020, liu_2020,kawabata_2021, liu_2020_b, leventis_2022, acharya_2022, weidemann_2021, longhi_2023_adp, ghatak_2024, tzortzakakis_2021, liu_2021_3, liu_2021_2, luo_2021, luo_2021_2, luo_2022, he_2024, sun_2024, li_2024, wang_2024, longhi_2024_arxiv, chakrabarty_2023, li_2024_arxiv}, driven by the synergy of Anderson localization and non-Hermiticity. This approach to non-Hermitian Anderson localization \cite{Sukhachov_2020,liu_2020,kawabata_2021, liu_2020_b, leventis_2022} allows for precisely controlled experiments, opening up a largely unexplored area of localization in complex media. Notable examples of intricate wave dynamics include constant-intensity waves despite strong localization \cite{makris_2017, makris_2020} and sudden Anderson jumps \cite{tzortzakakis_2021,weidemann_2021,leventis_2022,longhi_2023_adp, ghatak_2024} in disordered or quasiperiodic lattices. These concepts have been experimentally demonstrated in acoustics \cite{rivet_2018} and optics \cite{steinfurth_2022}, offering a different perspective in Anderson localization in open systems. 

However, wave propagation in conservative lattices is profoundly affected by their interaction with the environment. One possible way to effectively model such an interaction is by randomizing the wave function's phase, an effect known as dephasing. In periodic systems, such as lattices described by the Haken-Strobl-Reineker model \cite{kenkre_1982}, dephasing leads to dynamics similar to that of a Brownian particle \cite{moix_2013, madhukar_1977, witkoskie_2002, amir_2009}. In one- and two-dimensional disordered lattices, it is well-established that any non-zero level of disorder results in Anderson localization, thereby preventing transport. Nevertheless, studies examining dephasing, evolving disorder, or iterative measurements, suggest that the delicate interference essential to Anderson localization is disrupted, resulting in delocalization and incoherent, diffusive motion through classical inter-site hopping \cite{moix_2013, logan_1987, evensky_1990, schreiber_2011, gopalakrishnan_2017, rath_2020, flores_1999, gurvitz_2000, kosik_2006, kamiya_2015, broome_2010, znidaric_2013, yamada_1999, cao_2009, hoyer_2010, jayannavar_1988}. Very recently, the effect of dephasing on non-Hermitian systems \cite{longhi_2024_ol, longhi_2024_lsa} and Hermitian quasiperiodic systems  beyond the Anderson model \cite{bhakuni_2023, longhi_2024_prl} has started to gain attention, yet many aspects  remain largely unexplored.

This work aims to investigate the effects of dephasing on wave propagation in non-Hermitian lattices with complex on-site random disorder, where Anderson jumps can emerge under coherent dynamics \cite{leventis_2022, tzortzakakis_2021, weidemann_2021}. However, in the weak-disorder regime, these jumps are not apparent due to the large spatial extent of different eigenstates that are consequently overlapping. Counter-intuitively, in weakly disordered non-Hermitian lattices, rapid dephasing enhances eigenmode localization, resulting in jumpy wave function dynamics. We demonstrate that given the potential distributions and initial conditions, the position and duration of each jump can be analytically predicted \textit{a priori}. Additionally, we discuss the evolution of the derivative of optical power, which governs the system's dynamics, and the degree of localization of eigenmodes, as well as their dependence on the dephasing rate.


\section{Model under coherent and incoherent dynamics}

Our study begins with a one-dimensional (1D) non-Hermitian Anderson lattice consisting of $N$ evanescently coupled waveguides (indexed by $n \in \{ 1,2,...,N \}$)   that have complex on-site (gain and/or loss) strength. Thus, the Hamiltonian of the system is described by
\begin{equation}
    \hat{H} = \sum_{n=1}^{N-1} 
    \ket{n}\bra{n+1} + \ket{n+1}\bra{n} + \sum_{n=1}^{N}\epsilon_{n}\ket{n}\bra{n},
\end{equation}
where $\epsilon_{n} = a_{n} + ib_{n}\in \mathbb{C}$ takes random values from a rectangular distribution, with $a_{n} \in [-W_{R}/2, W_{R}/2]$ and $b_{n} \in [-W_{I}/2, W_{I}/2]$. We refer to $W_{R}, W_{I} \in \mathbb{R}^{+}$ as the real and imaginary disorder strengths, respectively. 

Under the paraxial approximation, the evolution of the wave function $\ket{\psi} \equiv \sum_{n=1}^{N} \psi_{n} \ket{n}$ in this lattice is described by the coupled-mode equation   
\begin{equation}
    \label{coherent}
    \hat{H}\ket{\psi} = -i \frac{d\ket{\psi}}{dz},
\end{equation}
where we assume that the waveguide channels form a linear chain with open boundary conditions (OBCs), i.e., $\psi_{0} = \psi_{N+1} = 0$.

The eigenstates of the right eigenvalue problem associated with this Hamiltonian are expressed as $\ket{\psi_{j}(z)} = \ket{u_{j}^{R}} e^{i\omega_{j}z}$, where $\ket{u_{j}^{R}}$ is one right eigenvector of the set $\{\ket{u_{m}^{R}}\}$, and $\omega_{j}$ is the corresponding eigenvalue, i.e., $\hat{H}\ket{u_{j}^{R}} = \omega_{j}\ket{u_{j}^{R}}$. The left eigenvalue problem associated with this Hamiltonian reads as $\hat{H}^{\dagger}\ket{u_{j}^{L}} = \omega_{j}^{*}\ket{u_{j}^{L}}$. The right and left eigenvectors of the Hamiltonian satisfy the biorthogonality condition $\bra{u_{k}^{L}}\ket{u_{j}^{R}} = \delta_{kj}$ while its symmetry ($\hat{H}=\hat{H}^{T}$) implies that $\ket{u_{j}^{L}}^{*}=\ket{u_{j}^{R}}$.
Therefore, $\ket{\psi}$ can be expressed as
\begin{equation}
    \label{psi_expansion}
    \ket{\psi(z)} = \sum_{j=1}^{N} 
    c_{j,0} e^{-\text{Im}(\omega_{j})z}e^{i\text{Re}(\omega_{j})z}
    \ket{u_{j}^{R}}\
\end{equation}
where $c_{j,0}\equiv \bra{u_{j}^{L}}\ket{\psi(0)}$, and $\ket{\psi(0)}= \sum_{n=1}^{N} \psi_{n}(0)\ket{n}$ is the initial condition. The non-Hermiticity of the Hamiltonian results in the non-conservation of optical power, defined as \(\mathcal{P}(z) \equiv \bra{\psi}\ket{\psi}\). To capture the wave packet's dynamics, we introduce a normalized wave function at each propagation distance \(z\), given by \(\ket{\phi} \equiv \ket{\psi}/\sqrt{\mathcal{P}(z)}\). This normalization can be implemented experimentally \cite{weidemann_2021} and ensures that the spatial profile of the wave function remains unchanged.

Regarding the dephasing, which is the focus of our study, we consider randomizing the phase of the wave function in a periodic fashion at each site $n$ at propagation distances $z_a = a \cdot l$, where $a \in \mathbb{N}$ and $l$ is the dephasing period. This is mathematically expressed as 
\begin{equation}
    \label{deph_schr}
    \psi_{n}(z_{a}^{+}) = e^{i \theta_{n}^{(a)}} \psi_{n}(z_{a}^{-}).
\end{equation}
where $\theta_{n}^{(a)} \in [0, 2\pi]$ are randomly selected phases. After statistical averaging over multiple realizations of the totally uncorrelated stochastic phases, the dynamical evolution of the average probability density \(P_{n}(z) \equiv \overline{|\psi_{n}(z)|^2}\) is described by
\begin{equation}
    \label{prop}
    P_{n}(z=al)=\mathcal{S}^{a}P_{n}(z=0),
\end{equation}
where \(\mathcal{S}(l)\) is the propagator matrix of the fully incoherent system, with elements given by \(\mathcal{S}_{kj}(l)=|(e^{iHl})_{kj}|^2\). Thus, the matrix $\mathcal{S}$, which we refer to as the incoherent propagator, is non-negative since $\mathcal{S}_{kj}\geq 0$. Since the Hamiltonian \( \hat{H} \) is symmetric (\( \hat{H} = \hat{H^{T}} \)), the propagator matrix \( U\equiv e^{iHl} \) is also symmetric, i.e., \( (e^{iHl})_{kj} = (e^{iHl})_{jk} \). Consequently, $\mathcal{S}$ is symmetric since \( \mathcal{S}_{kj} = |(e^{iHl})_{kj}|^{2} = |(e^{iHl})_{jk}|^{2} = \mathcal{S}_{jk} \). So, the eigenvalue equation for the symmetric and real-valued (Hermitian) matrix $\mathcal{S}$ is 
\begin{equation}
    \mathcal{S}\ket{v_{j}}=\beta_{j}\ket{v_{j}},
\end{equation}
From Eq.(~\ref{prop}), it follows that 
\begin{equation}
    \label{expansion}
        \ket{P(z=al)} = \sum_{j=1}^{N} d_{j,0} \beta_{j}^{a} \ket{v_{j}},
\end{equation}
where \(\ket{P(z)} \equiv \sum_{n=1}^{N} P_{n}(z) \ket{n}\) and \(d_{j,0} \equiv \bra{v_{j}}\ket{P(z=0)}\) [the derivation of Eqs. (\ref{prop}) and (\ref{expansion}) is provided in Appendix A].

In the case of \(W_{I} = 0\), where \(\hat{H}\) is Hermitian (\(\hat{H}^{\dagger} = \hat{H}\)), the propagator matrix \(U = e^{i\hat{H}l}\) is unitary (\(UU^{\dagger} = \hat{1}\)). Consequently, the relation $\sum_{k=1}^{N}\mathcal{S}_{kj}= \sum_{j=1}^{N}\mathcal{S}_{kj}=
1 $ for all rows and columns indexed by $k$ and $j$ holds, and since $\mathcal{S}$ is non-negative, it is by definition a doubly stochastic matrix. Thus, it can be directly shown that $\mathcal{S}$ has an eigenvalue \(\beta_{1} = 1\) corresponding to the eigenstate \(\ket{v_{1}} \sim (1,1,\ldots,1)^{T}\). All other eigenvalues of $\mathcal{S}$ have absolute values \(|\beta_{m}| \leq 1\), thus $\beta_{1}$ is the so-called Perron-Frobenius eigenvalue \cite{lawler2006}. Therefore, it follows from Eq.(~\ref{expansion}) that the projections of the average probability function onto eigenstates \(\{\ket{v_{m}}\}\) for $m\neq 1$ eventually vanish, and in the limit \(z \to \infty\), any initial excitation diffracts and adopts a homogeneous distribution in the lattice, i.e., \(\ket{P(z \to \infty)} = \ket{v_{1}}\).

For \(W_{I} > 0\), the non-unitarity of \(U\) results in a non-stochastic matrix \(\mathcal{S}\), thus lifting the constraint \(|\beta_{m}| \leq 1\), and the dynamical characteristics of the system change significantly, as discussed in Appendix B.
  
At this point, it is useful to introduce the inverse participation ratio (IPR), a widely used measure of wave function localization, which we use in the following sections. For the one-dimensional wave function $\ket{\xi}$, the IPR is defined as

\begin{equation}
    \text{IPR} \equiv \frac{\sum_{n=1}^{N} |\xi_{n}|^4}{\left(\sum_{n=1}^{N} |\xi_{n}|^2\right)^2}
\end{equation}
where $\xi_{n} \equiv \bra{n}\ket{\xi}$. This metric ranges from $1/N$ for a fully extended state, where $\xi_{n} = 1/\sqrt{N}$ for all $n$, to 1 for a completely localized state, where $\xi_{n} = \delta_{nm}$.

\section{Absence of jumps under coherent propagation at weak disorder}
Before further discussing the effect of dephasing, it is beneficial to review what happens under fully coherent conditions in a weakly non-Hermitian disordered lattice. In fact, a relevant phenomenon in such lattices, is the appearance of abrupt jumps between spatially distant regions of the lattice \cite{tzortzakakis_2021, leventis_2022, weidemann_2021}, as depicted in Fig. 1(a). These jumps occur between eigenstates that are localized in different regions and exhibit unequal rates of amplification or dissipation. 

The dynamics in such systems are governed by the evolution of the projection coefficients \( |c_j| = |c_{j,0}| e^{-\text{Im}(\omega_j)z} \) of Eq. (\ref{psi_expansion}). A jump occurs when the projection \( |c_k| \) of an eigenstate  \( \ket{u_{k}^{R}} \) surpasses the projection \( |c_r| \) of the previously dominant eigenstate \( \ket{u_{r}^{R}} \) (Fig. 1(b), where the condition \( \text{Im}(\omega_k) < \text{Im}(\omega_r) \) holds.  Ultimately, the eigenstate with the highest gain or the lowest loss (\( \min \{ \text{Im}(\omega_j) \} \)) dominates the system’s dynamics (Fig. 1(c)). 
However, these transitions are abrupt only in the high disorder regime, as lower disorder typically results in the presence of more extended eigenstates, which are spatially overlapping.

A comparison of the evolution of the normalized wave function \(\ket{\phi}\) for a single realization of high and low complex disorder, \(W_{R} = W_{I} = 6\) and \(W_{R} = W_{I} = 1\), respectively, in a lattice of \(N = 50\) waveguide channels under coherent dynamics (Eq. (\ref{coherent})) is shown in Fig. 1(a) (for strong disorder) and Fig. 1(d) (for low disorder). For both cases, we have considered a single-channel excitation at the middle of the lattice as initial condition, i.e., $\psi_{n}(0)=\delta_{n,25}$. In the high disorder case, the jump between distant regions of the lattice, seen in Fig. 1(a), corresponds to the crossing between dominant eigenstates, as indicated in Fig. 1(b). These eigenstates are highly localized, as evidenced by their large \(\text{IPR}\) (Fig. 1(c)). 

In the case of weak disorder, although there are two propagation distances at which transitions between dominant modes occur during the dynamics (Fig.~1(e)), no jumps are observed, as shown in Fig.~1(d). The absence of jumps arises because none of the three involved modes (nor any other eigenmodes of \(\hat{H}\)) are strongly localized. This is evidenced by their very low  (\text{IPR}) values, as illustrated in Fig.~1(f).

\begin{figure*}[!htbp]
	\includegraphics[width=1.0\textwidth]{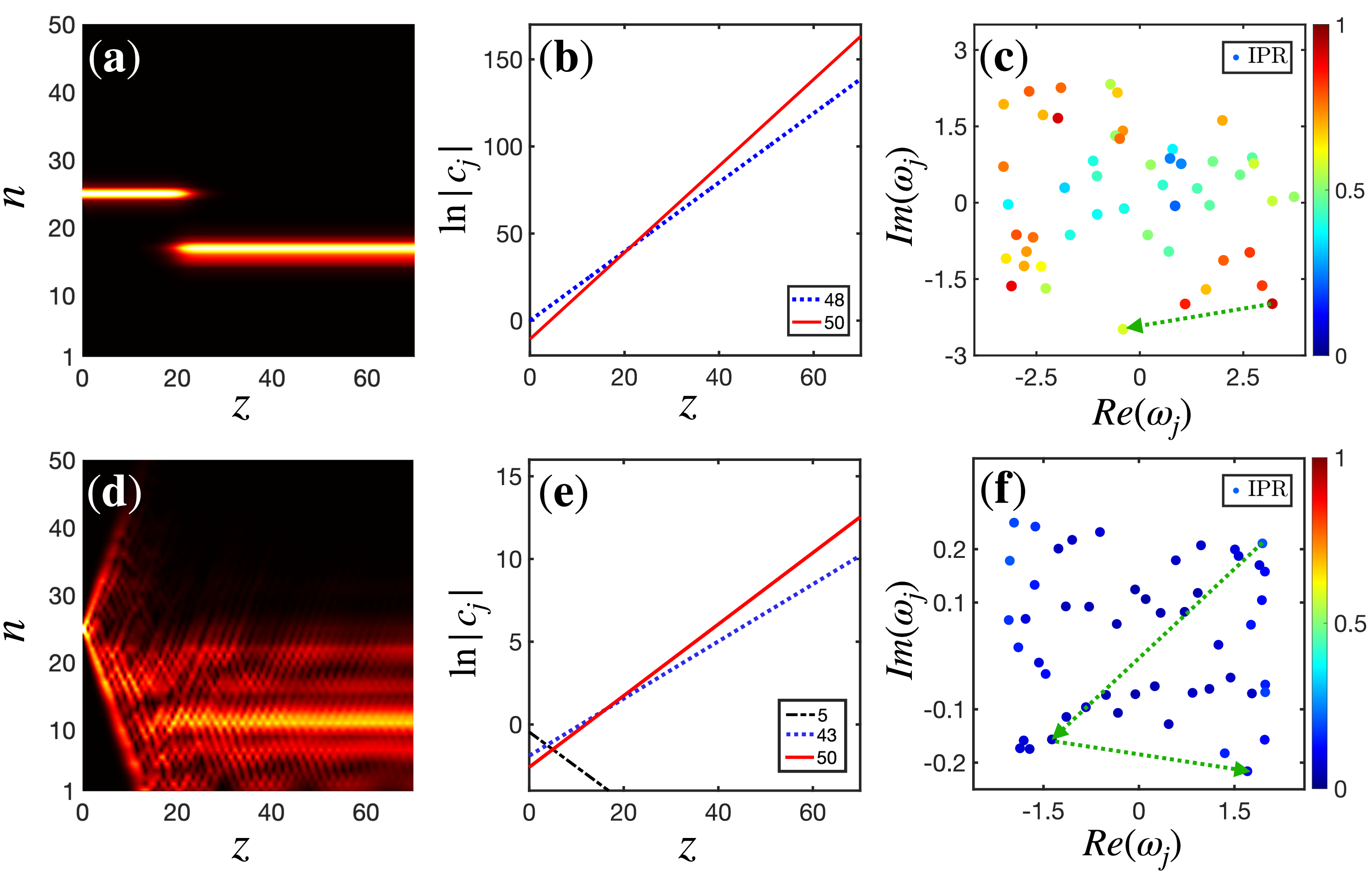}
	\caption{Comparison of coherent dynamics (Eq. (\ref{coherent}))for single realizations of strong disorder (\(W_{R}=W_{I}=6\), top row) and weak disorder (\(W_{R}=W_{I}=1\), bottom row) in lattices with \(N=50\) channels. Panels (a) and (d) show the evolution of the normalized wave function \(\ket{\phi}\) under single-channel excitation at \(n=25\) for the strong and weak disorder cases, respectively. Panels (b) and (e) show the logarithmic magnitudes of the projection coefficients \(\ln{|c_{j}|}\) for selected eigenstates, sorted by descending \(\text{Im}(\omega_{j})\), in the strong and weak disorder cases. Panels (c) and (f) show the eigenvalue spectrum of \(\hat{H}\) in the complex plane for the strong and weak disorder cases, respectively, with each eigenvalue \(\omega_{j}\) colored according to the inverse participation ratio (\(\text{IPR}\)) of the corresponding right eigenstate \(\ket{u_{j}^{R}}\). Arrows indicate the crossings between dominant projection coefficients as seen in panels (b) and (e).}
\end{figure*}

\section{EMERGENGE OF JUMPS DUE TO DEPHASING}

In the regime of fully incoherent dynamics, the evolution of the average probability density $\ket{P(z)}$ is described by Eq. (\ref{expansion}). Thus, the magnitude of the projections $d_{j} \equiv \bra{v_{j}}\ket{P(z)}$ of $\ket{P(z)}$ on the eigenstates $\ket{v_{j}}$ of the incoherent propagator $\mathcal{S}$ evolve as $|d_{j}|=|d_{j,0}|\beta_{j}^{z/l}$ so that
\begin{equation}
    \label{projections}
    \text{ln}|d_{j}|=
    \text{ln}|d_{j,0}|+\frac{\text{ln}\beta_{j}}{l}\cdot z
\end{equation}
From Eq. (\ref{projections}) it is evident that after a finite propagation distance $z_{cr}$ the eigenmode $\ket{v_{\max}}$ with the largest eigenvalue $\beta_{\max}\equiv\text{max}\{\beta_{j}\}$ will dominate the dynamics, giving the most significant contribution to the expansion of Eq. (\ref{expansion}).
Under fully incoherent dynamics, the average optical power of the system can be defined as
\begin{equation}
    \label{power}
    \overline{\mathcal{P}}(z)\equiv \sum_{n=1}^{N}
    P_{n}(z)\overset{(\ref{expansion})}{=}\sum_{n=1}^{N}\sum_{j=1}^{N}d_{j,0}v_{n,j}\beta_{j}^{z/l}
\end{equation}
where $v_{n,j}\equiv\bra{n}\ket{v_{j}}$, therefore, when one mode $\ket{v_{m}}$ dominates the dynamics, the above Eq. (\ref{power}) simplifies to
\begin{equation}
    \label{power_sim}
    \overline{\mathcal{P}}\approx
\beta_{m}^{z/l}\sum_{n=1}^{N}d_{m,0}v_{n,m}
\end{equation}
and it directly follows that 
\begin{equation}
    \label{power_der}
    \frac{d\ln{    \overline{\mathcal{P}}(z)}}{dz}=\frac{\ln{\beta_{m}}}{l}
\end{equation}
In Fig. 2(a), it is shown that for the same initial condition and disorder realization as in the bottom row of Fig. 1, the wave dynamics changes dramatically due to dephasing with a period of $l = 0.01$. This is evident from the abrupt transitions between highly localized regions, which, as shown in Fig. 2(b), occur at propagation distances corresponding to crossings between dominant projection coefficients $|d_{m}|$, whose evolution is described by Eq. (\ref{projections}). In Fig. 2(c), the real spectrum of $\mathcal{S}$ is illustrated, where as indicated by the high values of $\text{IPR}$, the corresponding eigenmodes $\ket{v_{j}}$, are highly localized. Thus, in the fully incoherent regime, the localization of the eigenmodes of $\mathcal{S}$ facilitates the occurrence of abrupt jumps. Additionally, as shown in Fig. 3(b), the derivative of the logarithm of optical power $\mathcal{P}(z)$ for a single realization of random phases matches the expected values from Eq. (\ref{power_der}). The latter are calculated based on the eigenvalue  $\beta_{m}$ corresponding to the dominant eigenmode $\ket{v_{m}}$ for each propagation interval.

It is important to note that the evolution of both the wave function $\ket{\phi}$ and the derivative of the optical power, $d\ln \mathcal{P}/dz$ [as shown in Fig. 2(a) and Fig. 3(b)], are numerically calculated for a single realization of random phases. Despite this, they are in good agreement with the theoretical predictions given by Eqs. (\ref{projections}) and (\ref{power_der}), respectively, although the aforementioned equations were derived under the assumption of statistical averaging.
\begin{figure*}[!htbp]
	\includegraphics[width=0.75\textwidth]{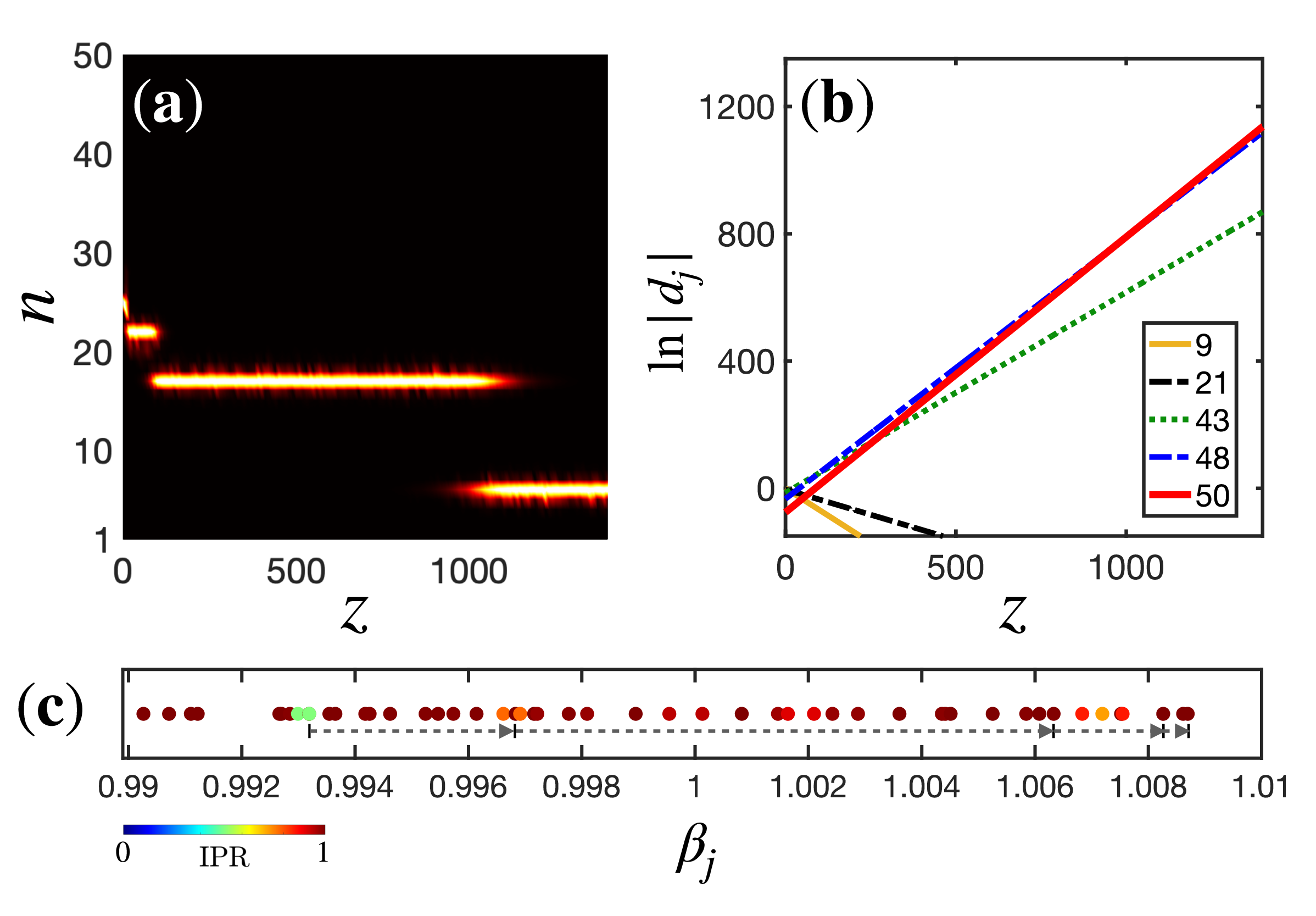}
	\caption{Single realization (same as the bottom row of  Fig. 1) of weak disorder (\(W_{R}=W_{I}=1\)) under fully incoherent dynamics [Eq. (\ref{deph_schr})], for a lattice of $N=50$ channels and a dephasing period of $l=0.01$. (a) Evolution of the normalized wave function $\ket{\phi}$ under single channel excitation at $n=25$, for a single random-phase realization (b) Logarithmic magnitudes of projection coefficients $\ln{|d_{j}|}$ for specific eigenstates, sorted in ascending values of $\beta_{j}$ (c) Eigenvalue spectrum of $\mathcal{S}$. Each point $\beta_{j}$ has a color associated with the $\text{IPR}$ of the corresponding  eigenstate $\ket{v_{j}}$. Arrows indicate the crossings between dominant projection coefficients as seen in panel (b).} 
\end{figure*}

\section{DEPENDENCE OF POWER DERIVATIVE AND DEGREE OF LOCALIZATION ON DEPHASING RATE}

Since $\hat{H}$ is non-Hermitian, the optical power $\mathcal{P}$ is  not conserved  under incoherent dynamics as well. In Fig. 3, we calculate the evolution of the derivative of the logarithm of optical power for the same single realization of weak complex disorder (\(W_{R}=W_{I}=1\)) as in Fig. 1 (bottom row) and Fig. 2, considering both coherent [Fig. 3(a)] and incoherent dynamics with rapid dephasing [\(l=0.01\), Fig. 3(b)]. It is clear that dephasing significantly changes the evolution of optical power. This observation naturally raises the question of whether and how this physical quantity depends on the dephasing period \(l\) in the fully incoherent regime. According to Eq. (\ref{power_der}), it is expected that beyond a finite distance \(z_{cr}(l)\), where the eigenmode \(\ket{u_{\text{max}}}\) becomes dominant, the average derivative of the optical power logarithm will remain constant. Consequently, this long-term power derivative depends on the dephasing period $l$ and the eigenvalue \(\beta_{\text{max}}\), which is itself a function of \(l\), according to Eq.  (\ref{power_der}). 
\begin{figure}[!htbp]
	\includegraphics[width=1\linewidth]{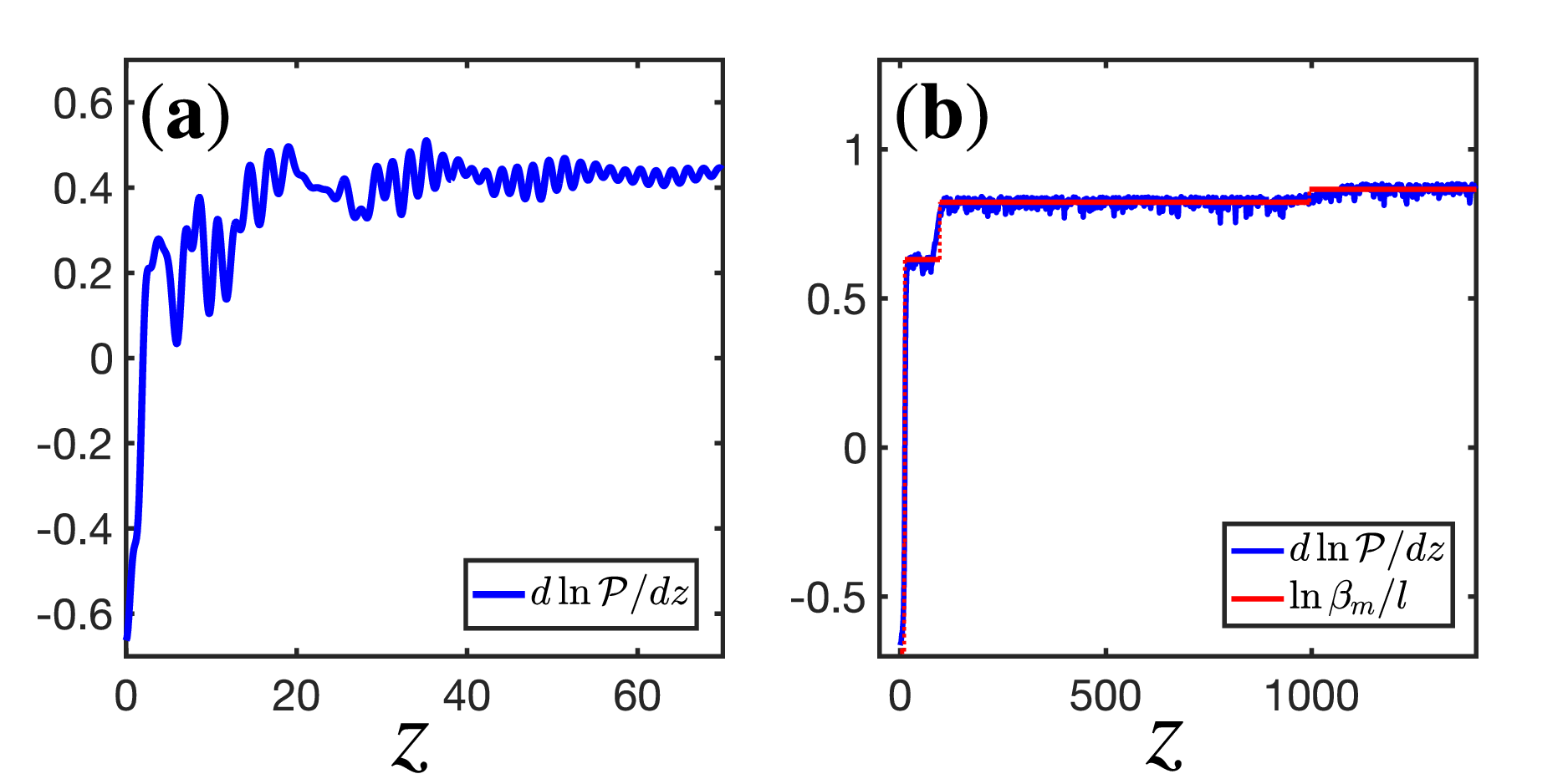}
	\label{aa}
	\caption{
Evolution of the derivative of the power logarithm for a single realization of weak disorder (\(W_{R}=W_{I}=1\)) corresponding to Fig. 1 (bottom row) and Fig. 2, under (a) coherent dynamics (no dephasing) and (b) incoherent dynamics with a dephasing period (\(l=0.01\)). In panel (b), the blue line represents the result from a single random-phase realization, while the red line indicates the expected values from Eq. (\ref{power_der}).
}

\end{figure}
As shown in Fig. 4(a), the long-term rate of the power derivative $d\ln{\overline{\mathcal{P}}(z>z_{cr})}/dz=\ln{\beta_{\max}}/{l}$ for the same single potential realization as Fig. 3, rises as the dephasing period $l$ decreases. This rate eventually saturates for a dephasing period of less than $l\sim10^{-3}$. 
\begin{figure}[!htbp]
	\includegraphics[width=0.5\textwidth]{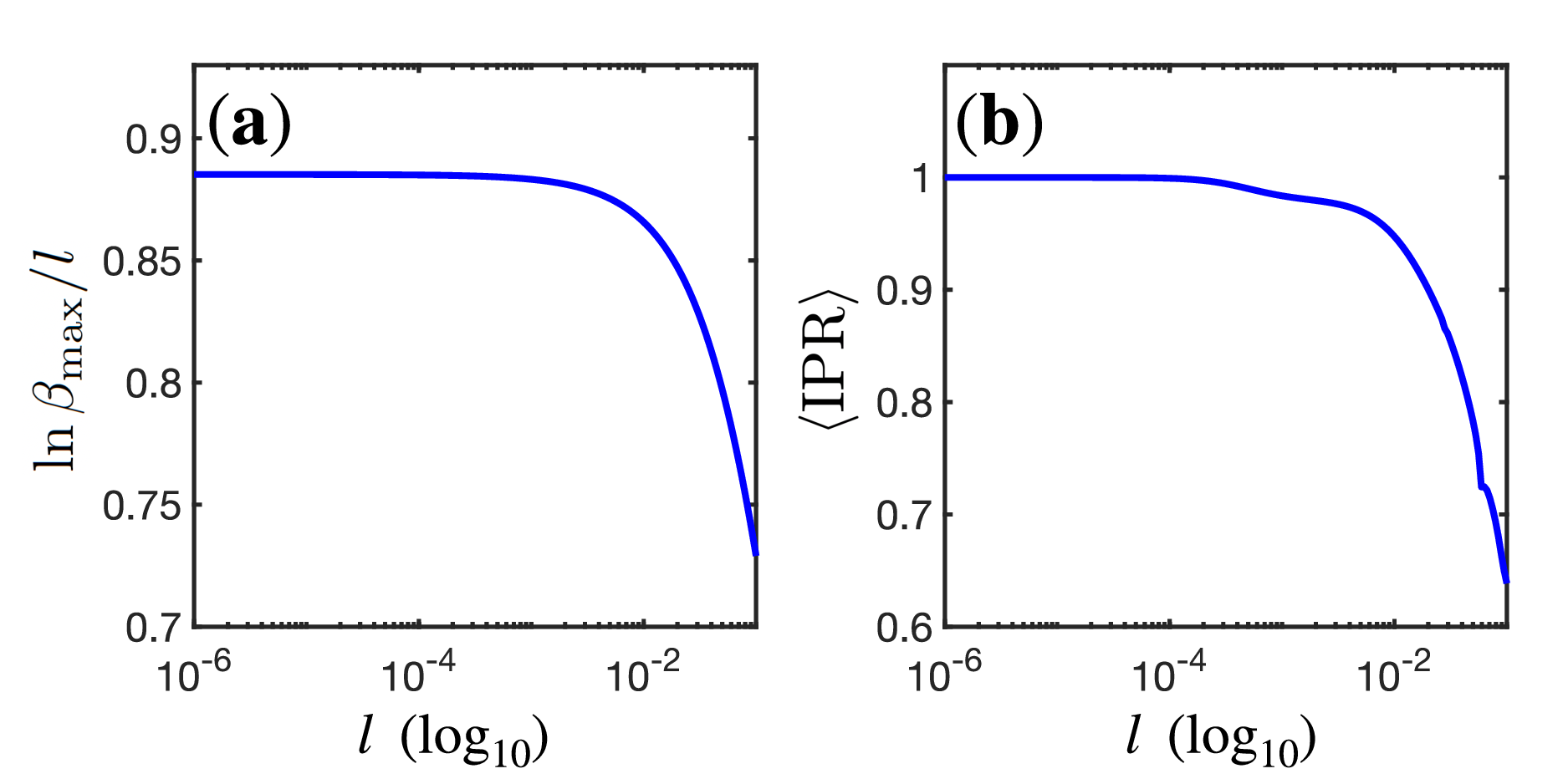}
	\caption{(a) Long-term average  power logarithm derivative as predicted by Eq. (\ref{power_der}) for the potential realization in Fig. 1 (bottom row) and Fig. 2. As \(l\) decreases, it saturates at \( d\ln{\overline{\mathcal{P}}(z)}/dz \approx 0.8852\). (b) Mean inverse participation ratio \(\langle \text{IPR} \rangle\) of the eigenmodes \(\ket{v_{j}}\) of the incoherent propagator \(\mathcal{S}(l)\), for the same potential realization. As \(l\) decreases, this saturates at \(\langle \text{IPR} \rangle = 1\), indicating localization to single sites.}
\end{figure}

It is also important to examine how the degree of localization of the eigenstates $\ket{v_{j}}$ of the incoherent propagator $\mathcal{S}$, as quantified by the inverse participation ratio $\text{IPR}$, depends on the dephasing period $l$. As shown in Fig. 4(b), using the same weak ($W_R=W_I=1$) disorder realization as in the previous figures, decreasing the dephasing period $l$ increases the mean inverse participation ratio $\langle \text{IPR} \rangle$ of the eigenstates $\ket{v_{j}}$, such that $\langle \text{IPR} \rangle \to 1$ for $l < 10^{-4}$. This indicates that below a certain value of the dephasing period $l$, all eigenstates of $\mathcal{S}$ become perfectly localized at single sites. Consequently, the incoherent dynamics of a single-channel excitation in such a lattice will primarily exhibit a jumpy evolution between distant single sites.

This behavior is not unique to the specific potential realization used in the previous analysis. As illustrated in Fig. 5, complete localization of the eigenmodes of the incoherent propagator occurs for all values of complex disorder strength $W \equiv W_{R} = W_{I}$, beyond the critical dephasing period $l_{cr}(W)$. For each value of disorder strength depicted in Fig. 5, 1000 different realizations of the random complex disorder were averaged to calculate the mean inverse participation ratio of the eigenmodes of $\mathcal{S}$, ensuring statistically reliable results.
\begin{figure}[!htbp]
	\includegraphics[width=1\linewidth]{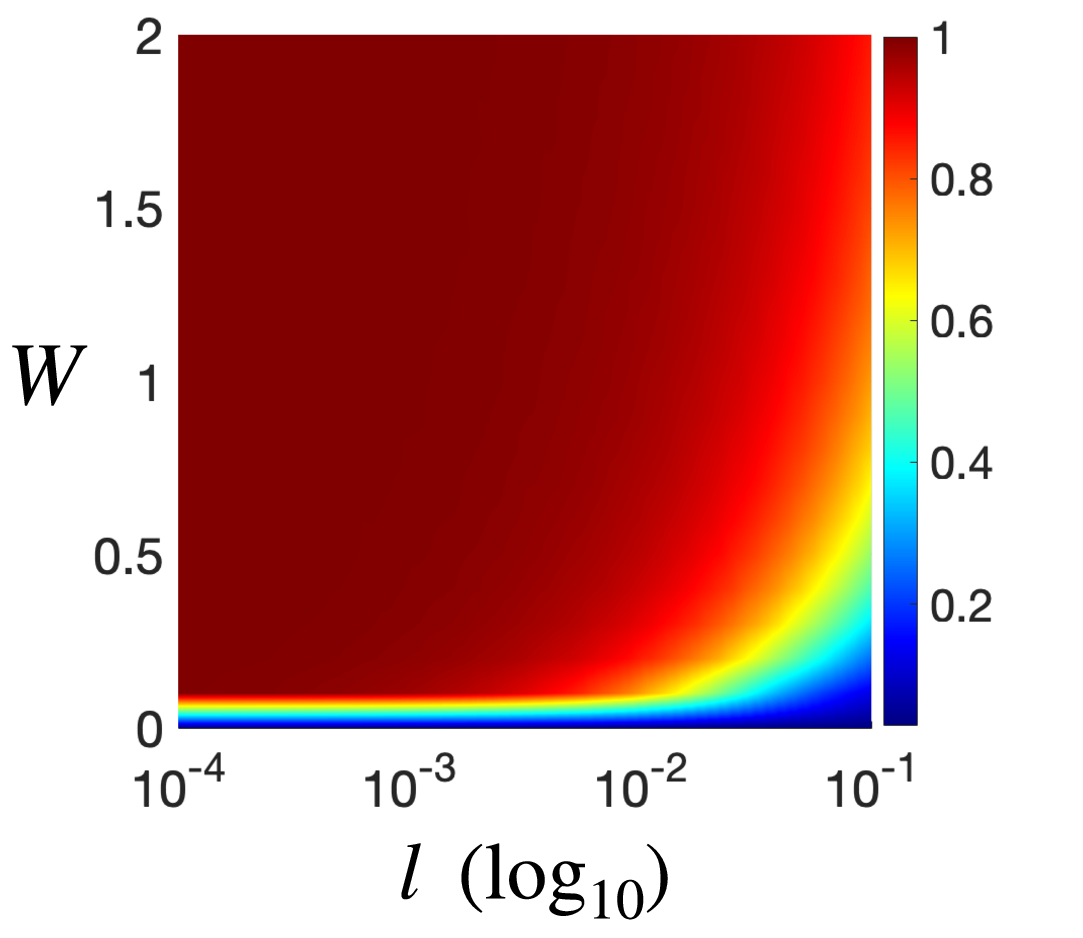}
	\label{deph_fig_5}
	\caption{Statistically averaged mean inverse participation ratio $\overline{\langle\text{IPR}\rangle}$ of the eigenmodes $\ket{v_{j}}$ of the incoherent propagator $\mathcal{S}(l)$, for various values of the dephasing period $l$ and the disorder strength $W_{R}=W_{I}\equiv W$. For each value of $W$, 1000 random realizations of potential were used for statistical averaging.}
\end{figure}
\section{Discussion and Conclusions}
In conclusion, we have investigated the impact of dephasing on wave propagation in non-Hermitian lattices with complex on-site weak disorder. 
Our analysis reveals that, while Anderson jumps are absent in the coherent case, dephasing, conversely, induces jumpy propagation. These phenomena should be experimentally observable in currently existing photonic discrete-time quantum walk platforms that use light-pulse propagation in synthetic-mesh lattices \cite{weidemann_2021}. Importantly, we demonstrate that the position and duration of each jump can be predicted analytically based on the given potential distributions and initial conditions. That comes in complete contradiction to the effect of dephasing on Hermitian disordered systems, where its presence destroys Anderson localization and leads to diffusion. Additionally, we examine the behavior of the optical power and show the saturation of its derivative with increasing dephasing frequency. The effect of the dephasing rate on eigenmode localization is also analyzed. These findings contribute to a deeper understanding of wave dynamics in non-Hermitian systems and are relevant for future theoretical and experimental studies in this area. 
\begin{acknowledgments}
The authors acknowledge the financial support provided by the European Research Council (ERC) under the Consolidator Grant Agreement No. 101045135 (Beyond$\_$Anderson). This research project was also co-funded by the Stavros Niarchos Foundation (SNF) and the Hellenic Foundation for Research and Innovation (H.F.R.I.) through the 5th Call of the “Science and Society” Action, titled “Always Strive for Excellence – Theodoros Papazoglou” (Project No: 11496, “PSEUDOTOPPOS”). 
\end{acknowledgments}

\appendix
\section{Incoherent propagation equation}

As also discussed in Ref. \cite{longhi_2024_lsa}, it follows from Eqs. (\ref{coherent}) and (\ref{deph_schr}), that the wave function's values at site $n$ between two consecutive dephasing events, i.e., $z_{a-1}=(a-1)l$ and $z_{a}=al$, where $a \in \mathbb{N}$, are connected through
\begin{equation}
    \psi_{n}\left(z_{a}\right)=
    e^{i\theta_{n}^{(a)}}
    \sum_{m=1}^{N} U_{n,m} \psi_{m}\left(z_{a-1}\right)
\end{equation}
and $U\equiv e^{iHl}$.
Recursively, the following equations hold
\begin{align}
    \psi_{n}\left(z_{a}\right) &=
    e^{i\theta_{n}^{(a)}}
    \sum_{m_{a}=1}^{N} U_{n,m_{a}} \psi_{m_{a}}\left(z_{a-1}\right) \notag \\
    \psi_{m_{a}}\left(z_{a-1}\right) &=
    e^{i\theta_{m_{a}}^{(a-1)}}
    \sum_{m_{a-1}=1}^{N} U_{m_{a},m_{a-1}} \psi_{m_{a-1}}\left(z_{a-2}\right) \notag \\
    &\vdots \notag \\
    \psi_{m_{2}}\left(z_{1}\right) &=
    e^{i\theta_{m_{2}}^{(1)}}
    \sum_{m_{1}=1}^{N} U_{m_{2},m_{1}} \psi_{m_{1}}\left(z_{0}\right)
\end{align}
so by consecutive substitutions one gets the expressions
\begin{equation}
    \begin{multlined}
        \psi_{n}\left(z_{a}\right) =
        \sum_{m_{1}, m_{2},..., m_{a}}^{N} 
        e^{i\theta_{n}^{(a)}+
        i\theta_{m_{a}}^{(a-1)}+...+i\theta_{m_{2}}^{(1)}}
        \times\\U_{n,m_{a}}U_{m_{a},m_{a-1}}... 
        U_{m_{2},m_{1}} \psi_{m_{1}}\left(z_{0}\right)
    \end{multlined}
\end{equation}
and
\begin{equation}
    \begin{multlined}
        \label{prob_sum}|\psi_{n}\left(z_{a}\right)|^2 =
        \sum_{m_{1}, \rho_{1},..., m_{a},\rho_{a}}^{N} 
        e^{i\theta_{m_{a}}^{(a-1)} - i\theta_{\rho_{a}}^{(a-1)}+...i\theta_{m_{2}}^{(1)} - i\theta_{\rho_{2}}^{(1)}}
        \times\\U_{n,m_{a}}U_{n,\rho_{a}}^{*}... U_{m_{2},m_{1}}U_{\rho_{2},\rho_{1}}^{*}\psi_{m_{1}}\left(z_{0}\right)\psi_{\rho_{1}}^{*}\left(z_{0}\right)
    \end{multlined}
\end{equation}
After statistical averaging over different realizations of the totally uncorrelated stochastic phases \(\theta_{n}^{(a)} \in [0, 2\pi]\), only terms with $m_{i}=\rho_{i}$ survive in the sum of Eq. (\ref{prob_sum}), so 
\begin{equation}
    \label{A6}
    P_{n}(z) \equiv \overline{|\psi_{n}(z)|^2} 
    = \sum_{m_{1},..., m_{a}}^{N}
       |U_{n,m_{a}}|^2 ... |U_{m_{2},m_{1}}|^{2} P_{m_{1}}\left(z_{0}\right)
\end{equation}
From this equation, it readily holds that
\begin{equation}
    P_{n}(z_{a})
    = \sum_{m=1}^{N}
       |U_{n,m}|^2 P_{m}\left(z_{a-1}\right)
\end{equation}
This difference equation describes the dynamics of any system in the totally incoherent regime and is mathematically equivalent to
\begin{equation}
    \psi_{n}(z_{a})
    = \sum_{m=1}^{N}
       U_{n,m} \psi_{m}\left(z_{a-1}\right)
\end{equation}
which is the well known expression for a system under coherent dynamics described by the Schr\"{o}dinger-type equation, Eq. (\ref{coherent}). Thus, in the fully incoherent regime, the proper propagator matrix $\mathcal{S}$ for the dynamics of the system has elements given by $\mathcal{S}_{nm}=|U_{n,m}|^2$. Recursively, from Eq. (\ref{A6}), we conclude to the expression
\begin{equation}
    \label{A8}
    P_{n}(z=al)=\mathcal{S}^{a}P_{n}(z=0),
\end{equation}
which is Eq. (\ref{prop}).
Applying Cayley-Hamilton's theorem to the eigenvalue equation of the incoherent propagator $\mathcal{S}$ we have
\begin{equation}
    \label{A10}\mathcal{S}^{a}\ket{v_{j}}=\beta_{j}^{a}\ket{v_{j}},
\end{equation}
We can express $\ket{P(z=0)}\equiv \sum_{n=1}^{N} P_{n}(z=0) \ket{n}$ in terms of the eigenstates of $\mathcal{S}$ as 
\begin{equation}
    \label{expansion_0}
    \ket{P(z=0)} = \sum_{j=1}^{N} d_{j,0}\ket{v_{j}},
\end{equation}
where \(d_{j,0} \equiv \bra{v_{j}}\ket{P(z=0)}\).
From Eqs. (\ref{A8}), (\ref{A10}), and (\ref{expansion_0}),  it follows that 
\begin{equation}
    \label{expansion_z}
    \ket{P(z=al)} = \sum_{j=1}^{N} d_{j,0} \beta_{j}^{a} \ket{v_{j}},
\end{equation}
which is Eq. (\ref{expansion}) of the main text.
{ 
\section{Localization of eigenstates of the incoherent propagator}
In the limit of fast dephasing, i.e., when \( l \ll 1 \), we can expand \( U \equiv e^{iHl}\) as \cite{longhi_2024_lsa}
\begin{equation}
    \label{exp}
    U \approx I + i\,{H}\,l - \frac{1}{2}\,{H}^2\,l^2,
\end{equation}
where $I$ is the identity matrix,
and we approximate the elements of \(\mathcal{S}_{kj} = U_{kj} U_{kj}^{*}\) up to the second order of $l$ as
\begin{equation}
\label{taylor}
{\small
\mathcal{S}_{kj} \approx
\begin{cases}
1 + i\Bigl( H_{kk} - H_{kk}^{*} \Bigr) l \\
\quad + \left( |H_{kk}|^2 - \frac{1}{2}\Bigl[ (H^2)_{kk} + (H^2)^*_{kk} \Bigr] \right) l^2, & k = j, \\[1ex]
|H_{kj}|^2\, l^2, & k = j\pm1.
\end{cases}
}
\end{equation} 
After straightforward calculations, one can show that for a Hamiltonian \(H\) with complex disordered on-site potentials, i.e., \(H_{nn}=a_{n}+ib_{n}\), the diagonal elements of \(\mathcal{S}\) are given by \(\mathcal{S}_{nn}\approx 1-2b_{n}l+(2b_n^2-2)l^2\) up to the second order in \(l\), while the off-diagonal elements are \(\mathcal{S}_{n,n\pm1}\approx l^2\). This effectively implies near-zero coupling between different sites, in the presence of fast dephasing. Thus, for \(l\ll1\) the eigenvalues of the incoherent propagator \(\mathcal{S}\) are given by $\beta_{n}\approx1-2b_{n}l+(2b_n^2-2)l^2$
up to the second order in \(l\), and consequently the eigenstates are highly localized in the lattice-site basis, i.e., \(\ket{v_n}\sim \ket{n}\).

From this discussion, it follows that the presence of imaginary on-site disorder (\(W_I>0\)) enhances the degree of localization of the eigenmodes and leads to jumpy dynamical evolution under dephasing. This localization persists regardless of the magnitude—or even the absence—of real disorder (\(W_R\)), since the real potentials \(a_n\) do not contribute to the eigenvalues of \(\mathcal{S}\) up to second order in \(l\). Consequently, dynamical delocalization through jumps can also be demonstrated in the presence of only imaginary disorder, and is solely caused by it.}


\end{document}